\documentclass[aps,a4paper,superscriptaddress,preprintnumbers,showpacs,
amsmath,amssymb]{revtex4}
\usepackage{graphicx}
\usepackage{dcolumn}
\usepackage{color}
\usepackage{latexsym,amsfonts}
\usepackage{textcomp}
\usepackage{bm}

\usepackage{ulem}

\begin{document}

\title{Quantum gravitational states\\ of ultracold neutrons as a tool
  for probing of beyond-Riemann gravity}

\author{A. N. Ivanov}\email{ivanov@kph.tuwien.ac.at}
\affiliation{Atominstitut, Technische Universit\"at Wien, Stadionallee
  2, A-1020 Wien, Austria}
\author{M. Wellenzohn}\email{max.wellenzohn@gmail.com}
\affiliation{Atominstitut, Technische Universit\"at Wien, Stadionallee
  2, A-1020 Wien, Austria} \affiliation{FH Campus Wien, University of
  Applied Sciences, Favoritenstra\ss e 226, 1100 Wien, Austria}
\author{H. Abele}\email{abele@ati.ac.at}
\affiliation{Atominstitut, Technische Universit\"at Wien, Stadionallee
  2, A-1020 Wien, Austria}

\date{\today}

\begin{abstract}
We analyze a possibility to probe beyond-Riemann gravity (BRG)
contributions, introduced by Kosteleck\'y and Li (see Phys. Rev. D
{\bf 103}, 024059 (2021) and Phys. Rev. D {\bf 104}, 044054 (2021)) on
the basis of the Effective Field Theory (EFT) by Kosteleck\'y
Phys. Rev. D {\bf 69}, 105009 (2004).  We carry out such an analysis
by calculating the BRG contributions to the transition frequencies of
the quantum gravitational states of ultracold neutrons (UCNs). These
states are being used for a test of interactions beyond the Standard
Model (SM) and General Relativity (GR) in the {\it q}BOUNCE
experiments. We improve by order of magnitude some constraints
obtained by Kosteleck\'y and Li (2106.11293 [gr-qc]).
\end{abstract}

\maketitle

\section{Introduction}
\label{sec:introduction}

Nowadays the effective field theories (EFT) are a powerful tool for
the analysis of the Nature \cite{Weinberg2021, Weinberg2016,
  Weinberg2009}. The general EFT by Kosteleck\'y
\cite{Kostelecky2004}, based on the General Gravity (GR) coupled to
the Standard Model (SM), has been extended by Kosteleck\'y and Li
\cite{Kostelecky2021a, Kostelecky2021b} by the contributions of
interactions, caused by beyond-Riemann gravity (BRG). These
contributions are closely overlapped with the contributions of
interactions violating Lorentz-invariance \cite{Kostelecky1997a,
  Kostelecky1998a, Kostelecky2002a, Kostelecky2002b, Kostelecky2003,
  Kostelecky2009, Kostelecky2016}. In \cite{Kostelecky2021b}
Kosteleck\'y and Li have proposed to investigate the BRG as well
as Lorentz-invariance violation (LV) contributions to the energy
spectrum and transition frequencies of the quantum gravitational
states of ultracold neutrons (UCNs).

This paper is addressed to the analysis of the BRG and LV
contributions to the energy spectrum and transition frequencies of the
quantum gravitational states of ultracold neutrons (UCNs). For the
analysis of these problems we follow \cite{Ivanov2019}. In
\cite{Ivanov2019} we have calculated the LV contributions to the
energy spectrum and transition frequencies of the quantum
gravitational states of UCNs, caused by the effective low-energy
potential, derived by Kosteleck\'y and Lane \cite{ Kostelecky1999} in
the frame work of the Standard Model Extension (SME)
\cite{Kostelecky1997a, Kostelecky1998a} (see also
\cite{Kostelecky2004}) by using the Foldy-Wouthuysen transformations
\cite{Foldy1950, Itzykson1980}.

The paper is organized as follows. In section \ref{sec:potential} we
discuss the effective low-energy potential, derived by Kosteleck\'y
and Li \cite{Kostelecky2021b} for the analysis of BRG interactions in
the terrestrial laboratories. We define such a potential in the
standard coordinate frame related to the laboratory at the Institut
Laue Langevin (ILL) in Grenoble. We specify the BRG and LV
contributions to the phenomenological coupling constants of this
potential. We adduce the wave functions of the quantum gravitational
states of polarized and unpolarized UCNs. In section \ref{sec:earth}
we calculate the BRG and LV contributions to the energy spectrum and
transition frequencies of the quantum gravitational states of
polarized and unpolarized UCNs. Using the current experimental
sensitivity of the {\it q}BOUNCE experiments we give some estimates of
the phenomenological constants of the BRG and LV interactions. In
section \ref{sec:Abschluss} we discuss the obtained results and
perspectives of further investigations of the BRG and LV interactions
by using the quantum gravitational states of UCNs.

\section{Effective non--relativistic potential of
  beyond-Riemann gravity interactions}
\label{sec:potential}

\begin{figure}
\includegraphics[height=0.26\textheight]{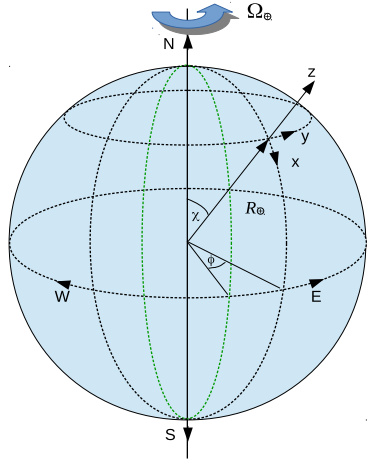}
  \caption{The position of the ILL laboratory of the {\it q}BOUNCE
    experiments on the surface of the Earth.}
\label{fig:fig1}
\end{figure}

For the experimental analysis of the BRG and LV interactions in the
terrestrial laboratories by using the quantum gravitational states of
UCNs Kosteleck\'y and Li propose to use the following Hamilton
operator \cite{Kostelecky2021b}
\begin{eqnarray}\label{eq:1}
  {\rm H} = {\rm H}_0 + \Phi_{\rm RG} + \Phi_{\rm BRG} =
  \frac{\vec{p}^{\,2}}{2m} - m \vec{g}\cdot \vec{z} + \Phi_{\rm nRG} +
  \Phi_{\rm nBRG},
\end{eqnarray}
where the first two terms are the operators of the UCN energy and the
Newtonian gravitational potential of the gravitational field of the
Earth, respectively, with the gravitational acceleration $\vec{g}$
such as $\vec{g}\cdot \vec{z} = - g z$ \cite{Kostelecky2021b}. Then,
$\Phi_{\rm nRG}$ is the effective low-energy potential of the
neutron-gravity interaction, calculated to next-to-leading order in
the large neutron mass $m$ expansion and related to the contribution
of Riemann gravity. It is equal to \cite{Kostelecky2021b}
\begin{eqnarray}\label{eq:2}
  \Phi_{\rm nRG} = \frac{3}{4m}\,\big(\vec{\sigma}\times
  \vec{p}\,\big)\cdot \vec{g} -
  \frac{3}{4m}\,\big(\vec{p}^{\,2}\,\vec{g}\cdot \vec{z} +
  \vec{g}\cdot \vec{z}\,\vec{p}^{\,2}\big).
\end{eqnarray}
In turn, the potential $\Phi_{\rm nBRG}$ describes the BRG and LV
contributions to neutron-gravity interactions
\begin{eqnarray}\label{eq:3}
  \Phi_{\rm nBRG} = H_{\phi} + H_{\sigma \phi} + H_g + H_{\sigma g},
\end{eqnarray}
where the operators $H_j$ for $j = \phi, \sigma\phi, g$ and $ \sigma
g$ are equal to \cite{Kostelecky2021b}
\begin{eqnarray}\label{eq:4}
 H_{\phi} &=& (k^{\rm NR}_{\phi})_n \vec{g}\cdot \vec{z} + (k^{\rm
   NR}_{\phi p})^j_n \frac{1}{2}\Big(p^j(\vec{g}\cdot \vec{z}\,) +
 (\vec{g}\cdot \vec{z}\,) p^j\Big) + (k^{\rm NR}_{\phi pp})^{jk}_n
 \frac{1}{2}\Big(p^jp^k (\vec{g}\cdot \vec{z}\,) + (\vec{g}\cdot
 \vec{z}\,) p^jp^k\Big),\nonumber\\ H_{\sigma\phi} &=&(k^{\rm
   NR}_{\sigma \phi})^j_n \sigma^j(\vec{g}\cdot \vec{z}\,) +
 (k^{(NR)}_{\sigma \phi p})^{jk}_n \frac{1}{2}\,\sigma^j
 \Big(p^k(\vec{g}\cdot \vec{z}\,) + (\vec{g}\cdot \vec{z}\,) p^k\Big)
 + (k^{\rm NR}_{\sigma\phi pp})^{jk\ell}_n
 \frac{1}{2}\,\sigma^j\Big(p^kp^{\ell} (\vec{g}\cdot \vec{z}\,) +
 (\vec{g}\cdot \vec{z}\,) p^kp^{\ell}\Big),\nonumber\\ H_g &=&
 (k^{(NR)}_g)^j_n g^j + (k^{(NR)}_{g p})^{jk}_n p^j g^k + (k^{(NR)}_{g
   pp})^{jk\ell}_n p^j p^k g^{\ell},\nonumber\\ H_{\sigma g} &=&
 (k^{\rm NR}_{\sigma g})^{jk}_n \sigma^j g^k + (k^{\rm NR}_{\sigma g
   p})^{jk\ell}_n \sigma^j p^k g^{\ell} + (k^{\rm NR}_{\sigma g
   pp})^{jk\ell m}_n \sigma^j p^kp^{\ell} g^m.
\end{eqnarray}
The non-relativistic Hamilton operator Eq.(\ref{eq:1}) is written in
the coordinate system shown in Fig.\,\ref{fig:fig1}, where $m$ is the
neutron mass, $\vec{z}$ is a radius-vector of a position of an UCN on
the $z$-axis, $\vec{p} = - i \nabla$ is a 3-momentum of an UCN and
$\vec{\sigma}$ is the Pauli $2 \times 2$ matrix of the UCN spin
\cite{Itzykson1980}. The coefficients $(k^{\rm NR}_{\phi})_n$,
$(k^{\rm NR}_{\ldots})^j_n$, $(k^{\rm NR}_{\ldots})^{jk}_n$, $(k^{\rm
  NR}_{\ldots})^{jk\ell}_n$, and $(k^{\rm NR}_{\ldots})^{jk\ell m}_n$
define the BRG and LV contributions, which can be tested in
experiments with neutrons \cite{Kostelecky2021b} in the following way.

The system of a Schrödinger quantum particle with mass $m$ bouncing in
a linear gravitational field is known as the quantum bouncer
\cite{Gibbs1975, Gea-Banacloche1999, Rosu2001, Robinett2004}. Above a
horizontal mirror, the linear gravity potential leads to discrete
energy eigenstates of a bouncing quantum particle. An UCN, bound on a
reflecting mirror in the gravity potential of the earth, can be found
in a superposition of quantum gravitational energy eigen-states. The
quantum gravitational states of UCNs have been verified and
investigated \cite{Nesvizhevsky2002, Nesvizhevsky2003,
  Nesvizhevsky2005, Abele2007} at the UCN beamline PF2 at the
Institute Laue-Langevin (ILL), where the highest UCN flux is available
worldwide.  The {\it q}BOUNCE collaboration develops a gravitational
resonant spectroscopy (GRS) method \cite{Abele2010}, which allows to
measure the energy difference between quantum gravitational states
with increasing accuracy. Recent activities \cite{Sedmik2019}, and a
summary can be found in \cite{Jenke2019}. The energy difference can be
related to the frequency of a mechanical modulator, in analogy to the
Nuclear Magnetic Resonance technique, where the Zeeman energy
splitting of a magnetic moment in an outer magnetic field is connected
to the frequency of a radio-frequency field. The frequency range in
GRS used so far is in the acoustic frequency range between 100 and
1000\,${\rm Hz}$. The quantum gravitational states of UCNs have peV
energy, on a much lower energy scale compared to other bound quantum
systems. Any gravity-like potential or a deviation from Riemann
gravity would shift these energy levels \cite{Jenke2021, Jenke2011,
  Jenke2014a, Cronenberg2018} and an observation would point to new
physical understanding.

Our choice of the laboratory frame is related to the
following. Indeed, the {\it q}BOUNCE experiments are being performed
in the laboratory at Institut Laue Langevin (ILL) in Grenoble. The ILL
laboratory is fixed to the surface of the Earth in the northern
hemisphere. Following \cite{Kostelecky2002a, Kostelecky2002b,
  Kostelecky2003, Kostelecky2009, Kostelecky2016} (see also
\cite{Kostelecky2021b, Ivanov2019}) we choose the ILL laboratory or
the standard laboratory frame with coordinates $(t, x, y, z)$, where
the $x$, $y$ and $z$ axes point south, east and vertically upwards,
respectively, with northern and southern poles on the axis of the
Earth's rotation with the Earth's sidereal frequency $\Omega_{\oplus}
= 2\pi/(23\,{\rm hr}\, 56\,{\rm min}\, 4.09\,{\rm s} = 7.2921159
\times 10^{-5}\,{\rm rad/s}$. The position of the ILL laboratory on
the surface of the Earth is determined by the angles $\chi$ and
$\phi$, where $\chi = 44.83333^0$\,N is the colatitude of the
laboratory and $\phi$ is the longitude of the laboratory measured to
east with the value $\phi = 5.71667^0$\,E \cite{Grenoble}. The beam of
UCNs moves from south to north antiparallel to the $x$--direction and
with energies of UCNs quantized in the $z$--direction. The
gravitational acceleration in Grenoble is $g = 9.80507\,{\rm m/s^2}$
\cite{Ivanov2019, Grenoble}. Following \cite{Ivanov2019} we may
neglect the Earth's rotation assuming that the ILL laboratory frame is
an inertial one.

\subsection*{\bf The analysis of 
  contributions to the effective low-energy potential $\Phi_{\rm
    nBRG}$ in Eq.(\ref{eq:3}) violating of Lorentz-invariance}

Before we proceed to calculating the contributions of the effective
potential Eq.(\ref{eq:3}) to the energy spectrum and transition
frequencies of the quantum gravitational states of UCNs we would like
to compare the potential $\Phi_{\rm nBRG}$ with the effective
low-energy potential $\Phi_{\rm nLV}$ of the LV interactions (see
Eq.(\ref{eq:4}) in \cite{Ivanov2019}), calculated in
\cite{Kostelecky1999}. The effective low-energy potential $\Phi_{\rm
  nLV}$ is equal to
\begin{eqnarray}\label{eq:5}
\hspace{-0.15in}\Phi_{\rm nLV} &=& \Big(- b^n_{\ell} + m d^n_{\ell 0}
- \frac{1}{2}\,m \,\varepsilon_{\ell k j} g^n_{k j 0} +
\frac{1}{2}\,\varepsilon_{\ell k j} H^n_{k j}\Big) \sigma_{\ell} +
\frac{1}{m}\, \Big(- a^n_j + m(c^n_{0j} + c^n_{j0}) + m e^n_j\Big)
p_j\nonumber\\
\hspace{-0.15in}&+& \frac{1}{m}\, \Big(b^n_0 \delta_{j\ell} -
m(d^n_{\ell j} + d^n_{00} \delta_{\ell j}) - \frac{1}{2}\,m\,
\varepsilon_{\ell k m}\big(g^n_{m k j} + 2 g^n_{m00}\delta_{j k}\big)
- \varepsilon_{j \ell k} H^n_{k 0}\Big) p_j \sigma_{\ell} - \frac{1}{2
  m}\,\Big(2 c^n_{jk} + c^n_{00}\delta_{jk}\Big) p_j p_k \nonumber\\
\hspace{-0.15in}&+& \Big[\frac{1}{4 m}\,\Big(\big(4 d^n_{0j} + 2
  d^n_{j0} - \varepsilon_{j m n} g^n_{m n 0}\big)\, \delta_{k\ell} +
  \varepsilon_{\ell m n} g^n_{mn0}\, \delta_{jk} - 2\,\varepsilon_{j
    \ell m}\,\big(g^n_{m 0 k} + g^n_{m k 0}\big)\Big)
  \nonumber\\\hspace{-0.15in}&+& \frac{1}{2 m^2}\,\Big(\big( - b_j -
  \frac{1}{2}\,\varepsilon_{j m n} H_{mn}\big)\,\delta_{k\ell} +
  b_{\ell} \delta_{jk}\Big)\Big]\,p_j p_k \sigma_{\ell}.
\end{eqnarray}
The LV contributions to the energy spectrum and transition frequencies
of the quantum gravitational states of UCNs, induced by the effective
low-energy potential Eq.(\ref{eq:5}), have been calculated in
\cite{Ivanov2019}.

From Eq.(\ref{eq:4}) one may see that the effective low-energy
interactions $H_{\phi}$, $H_{\sigma \phi}$ and $(k^{(NR)}_g)^j_n g^j$
in $H_g$ are new in comparison with Eq.(\ref{eq:5}). So this means
that the coefficients or the phenomenological coupling constants in
these interactions are induced by the BRG interactions. Of course,
these terms are able to contain the LV contributions (see Table III of
Ref.\cite{Kostelecky2021b}) but such contributions should not dominate
in them.

In turn, the effective low-energy neutron-gravity interactions,
defined by $H_g$ and $H_{\sigma g}$, have the structure of the
effective low-energy potential $\Phi_{\rm nLV}$ in
Eq.(\ref{eq:5}). From the comparison we may write the following
relations
\begin{eqnarray}\label{eq:6}
(k^{\rm NR}_{\sigma g})^{jk}_ng^k &=& - b^n_j + m d^n_{j 0} -
  \frac{1}{2}\,m \,\varepsilon_{j k \ell} g^n_{k \ell 0} +
  \frac{1}{2}\,\varepsilon_{j k \ell} H^n_{k \ell} +
  \ldots,\nonumber\\ (k^{\rm NR}_{g p})^{jk}_ng^k &=&\frac{1}{m}\, \Big(-
  a^n_j + m(c^n_{0j} + c^n_{j0}) + m e^n_j\Big) +
  \ldots,\nonumber\\ (k^{\rm NR}_{\sigma g p})^{jk\ell}_ng^{\ell}
  &=&\frac{1}{m}\, \Big(b^n_0 \delta_{j k} - m(d^n_{k j} + d^n_{00}
  \delta_{k j}) - \frac{1}{2}\,m\, \varepsilon_{j t m}\big(g^n_{m t k}
  + 2 g^n_{m00}\delta_{j k}\big) - \varepsilon_{j k m} H^n_{m 0}\Big)
  + \ldots,\nonumber\\ (k^{\rm NR}_{g p p})^{jk\ell}_ng^{\ell} &=& -
  \frac{1}{2 m}\,\Big(2 c^n_{jk} + c^n_{00}\delta_{jk}\Big) +
  \ldots,\nonumber\\ (k^{\rm NR}_{\sigma g p p})^{jk\ell m}_ng^m &=&
  \frac{1}{4 m}\,\Big(\big(4 d^n_{0j} + 2 d^n_{j0} - \varepsilon_{j m
    n} g^n_{m n 0}\big)\, \delta_{k\ell} + \varepsilon_{\ell m n}
  g^n_{mn0}\, \delta_{jk} - 2\,\varepsilon_{j \ell m}\,\big(g^n_{m 0
    k} + g^n_{m k 0}\big)\Big)\nonumber\\ &+& \frac{1}{2
    m^2}\,\Big(\big( - b_j - \frac{1}{2}\,\varepsilon_{j m n}
  H_{mn}\big)\,\delta_{k\ell} + b_{\ell}\delta_{jk}\Big) + \ldots,
\end{eqnarray}
where ellipses denote the BRG contributions of neutron-gravity
interactions (see Table III in Ref.\cite{Kostelecky2021b}).

\subsection*{\bf The rotation-invariant effective low-energy potential
  $\Phi^{(\rm RI)}_{\rm nBRG}$ for the {\it q}BOUNCE experiments}

For the experimental analysis of the BRG as well as LV interactions by
the quantum gravitational states of UCNs Kosteleck\'y and Li proposed
to use the following rotation-invariant (RI) effective low-energy
potential \cite{Kostelecky2021b}
\begin{eqnarray}\label{eq:7}
  \Phi^{(\rm RI)}_{\rm BRG} &=& \big(k^{\rm NR}_{\phi})_n\,
  \vec{g}\cdot \vec{z} + \big(k^{\rm NR}_{\sigma
    g}\big)'_n\,\vec{\sigma}\cdot \vec{g} + \big(k^{\rm NR}_{\sigma g
    \phi}\big)'_n\,\big(\vec{\sigma}\times \vec{p}\,\big)\cdot \vec{g}
  + \frac{1}{2}\,\big(k^{\rm NR}_{\sigma \phi
    p}\big)'_n\,\Big((\vec{\sigma}\cdot \vec{p}\,)(\vec{g}\cdot
  \vec{z}\,) + (\vec{g}\cdot \vec{z}\,)(\vec{\sigma}\cdot
  \vec{p}\,)\Big)\nonumber\\ &+& \big(k^{\rm NR}_{\sigma g
    pp}\big)'_n\,\vec{p}^{\,2}\,\vec{\sigma}\cdot \vec{g} +
  \big(k^{\rm NR}_{\sigma g pp}\big)''_n\,
  (\vec{\sigma}\cdot\vec{p}\,)(\vec{g}\cdot \vec{p}\,).
\end{eqnarray}
In this expression the coefficients with primes denote suitably
normalized irreducible representations of the rotation group obtained
from the nonrelativistic coefficients in Eq.(\ref{eq:4}) (see
\cite{Kostelecky2021b}). Then, according to Kostelecky and Li
\cite{Kostelecky2021b}, the effective low-energy potential
Eq.(\ref{eq:4}) is of interest for certain experimental applications,
in part because the rotation invariance ensures that all terms take
the same form at leading order when expressed either in the laboratory
frame or the Sun-centered frame. The latter implies, for example, no
leading-order dependence on the local sidereal time or laboratory
colatitude in experimental signals for these terms
\cite{Kostelecky2021b}.

Since the effective neutron-gravity interactions, proportional to $
\big(k^{(\rm NR)}_{\sigma g \phi}\big)'_n$ and $\big(k^{(\rm
  NR)}_{\sigma \phi p}\big)'_n$, do not contribute to the energy
spectrum of the quantum gravitational states of UCNs, the possible
contributions should be proportional to the coefficients $ \big(k^{\rm
  NR}_{\phi})_n$, $ \big(k^{\rm NR}_{\sigma g}\big)'_n$, $\big(k^{\rm
  NR}_{\sigma g pp}\big)'_n$ and $\big(k^{\rm NR}_{\sigma g
  pp}\big)''_n$, respectively. According to our discussion above, the
coefficient $ \big(k^{\rm NR}_{\phi})_n$ is caused by the BRG
interactions, whereas the coefficients $ \big(k^{\rm NR}_{\sigma
  g}\big)'_n$, $\big(k^{\rm NR}_{\sigma g pp}\big)'_n$ and
$\big(k^{\rm NR}_{\sigma g pp}\big)''_n$ should be saturated by the LV
ones \cite{Ivanov2019}.

\subsection*{\bf Wave functions and energy spectrum of quantum
  gravitational states of UCNs}

The non-perturbed quantum gravitational states of UCNs obey the
Schr\"odinger-Pauli equation \cite{Gibbs1975}
\begin{eqnarray}\label{eq:8}
 i\frac{\partial \Psi^{(0)}_{\vec{p}_{\perp} k
     \sigma}(t,\vec{r}\,)}{\partial t} = \Big(\frac{\vec{p}^{\,2}}{2
   m} + m g z\Big)\Psi^{(0)}_{\vec{p}_{\perp} k \sigma}(t,\vec{r}\,)
\end{eqnarray}
where $\vec{r} = \vec{z} + \vec{r}_{\perp}$ is a radius-vector of a
position of an UCN with $\vec{r}_{\perp} =(x,y)$, the wave function
$\Psi_{\vec{p}_{\perp} k\sigma}(t,\vec{r}\,)$ is equal to
$\Psi_{\vec{p}_{\perp} k\sigma}(t,\vec{r}\,) =
\Psi^{(0)}_{k\sigma}(z)\, e^{\,i \vec{p}_{\perp}\cdot \vec{r}_{\perp}
  - i(E_{\vec{p}} + E^{(0)}_k)t}/2\pi = |\vec{p}_{\perp} k
\sigma\rangle$, $\vec{p}_{\perp}$ and $E_{\vec{p}_{\perp}} =
\vec{p}^{\,2}_{\perp}/2m \sim 10^{-7}\,{\rm eV}$ (for
$|\vec{p}_{\perp}| \sim 24\,{\rm eV}$ or $v_{\perp} \sim 7\,{\rm
  m/s}$) are the momentum and kinetic energy of UCNs. Below all BRG
and LV contributions will be calculated at $\vec{p}_{\perp} = 0$
\cite{Ivanov2019}, and $k = 1,2,\ldots$ is the principal quantum
number \cite{Gibbs1975}. The wave function $\Psi_{\vec{p}_{\perp}
  k\sigma}(t,\vec{r}\,)$ is normalized by $\langle \sigma' k'
\vec{p}^{\,'}_{\perp}|\vec{p}_{\perp} k \sigma\rangle =
\delta^{(2)}(\vec{p}^{\,'}_{\perp} -
\vec{p}_{\perp})\,\delta_{k'k}\delta_{\sigma'\sigma}$ \cite{LL1965,
  Davydov1965}. Then, $\Psi^{(0)}_{k\sigma}(z) =
\psi^{(0)}_k(z)\,\chi_{\sigma} = |k\sigma\rangle$ is a two--component
spinorial wave function of UCNs in the $k$--gravitational state with
the binding energy $E^{(0)}_k$, and in a spin eigenstate
$\chi_{\sigma}$ with $\sigma = \uparrow$ or $\downarrow$. They are
normalized by $\langle \sigma'k'|k\sigma\rangle =
\delta_{k'k}\delta_{\sigma'\sigma}$. The wave functions
$\psi^{(0)}_k(z)$ are given by \cite{Gibbs1975}
\begin{eqnarray}\label{eq:9}
 \psi^{(0)}_k(z) = \frac{\displaystyle {\rm Ai}(\xi -
   \xi_k)}{\sqrt{\ell}\,|{\rm Ai}'(-\xi_k)|}\,e^{\,i\alpha}
 \quad,\quad \int^{\infty}_0dz\,\psi^{(0)*}_{k'}(z)\psi^{(0)}_k(z) =
 \delta_{k'k},
\end{eqnarray}
 where $\xi = z/\ell$, ${\rm Ai}(\xi - \xi_k)$ is the Airy-function
 and ${\rm Ai}'(-\xi_k)$ its derivative at $z = 0$ \cite{Gibbs1975,
   HMF72, Albright1977, AiryF2004}, $e^{\,i\,\alpha}$ is a constant
 complex factor, $\ell = (2m^2g)^{-1/3} = 5.88\,{\rm \mu m}$ is the
 scale of the quantum gravitational states of UCNs and $\xi_k$ is the
 root of the equation ${\rm Ai}(-\xi_k) = 0$, cased by the boundary
 condition $\psi^{(0)}_k(0) = 0$ \cite{Gibbs1975}. The latter defines
 the energy spectrum of the quantum gravitational states of UCNs
 $E^{(0)}_k = E_0\,\xi_k$ for $k = 1,2,\ldots$ with $E_0 = m g \ell =
 \sqrt[3]{m g^2/2} = 0.6016\,{\rm peV}$
 \cite{Ivanov2019}. Experimentally the quantum gravitational states of
 UCNs have been investigated in \cite{Nesvizhevsky2002,
   Nesvizhevsky2003, Nesvizhevsky2005}.

 The wave functions $\Psi^{(0)}_{k\sigma}(z) =
 \psi^{(0)}_k(z)\,\chi_{\sigma}$ describe the quantum gravitational
 states of polarized UCNs, whereas for the quantum gravitational
 states of unpolarized UCNs the wave functions are given by
 \cite{Ivanov2019}
 \begin{eqnarray}\label{eq:10}
\Psi^{(0)}_k(z) = \psi^{(0)}_k(z)\,c_{\uparrow}\,\chi_{\uparrow} +
\psi^{(0)}_k(z)\,c_{\downarrow}\,\chi_{\downarrow},
\end{eqnarray}
where the coefficients $c_{\uparrow}$ and $c_{\downarrow}$ are
normalized by $|c_{\uparrow}|^2 + |c_{\downarrow}|^2 = 1$ and
determine the probabilities to find an UCN in the $k$--quantum
gravitational state with spin {\it up} and {\it down},
respectively. The quantum gravitational states of UCNs with the wave
function Eq.(\ref{eq:6}) are 2--fold degenerate \cite{LL1965,
  Davydov1965}.

\section{The BRG and LV contributions to the energy spectrum and
  transition frequencies of quantum gravitational states of UCNs}
\label{sec:earth}

The energy spectrum of the quantum gravitational states of polarized
UCNs with the RG, BRG and LV corrections are defined by the integrals
\begin{eqnarray}\label{eq:11}
E_{k \sigma} = \langle \sigma k|{\rm H}|k \sigma\rangle =
\int^{\infty}_0 dz\,\Psi^{(0)\dagger}_{k\sigma}(z){\rm
  H}\Psi^{(0)}_{k\sigma}(z) = E^{(0)}_k + \langle \sigma k|\Phi_{\rm
  nRG}|k\sigma\rangle + \langle \sigma k|\Phi^{(\rm RI)}_{\rm
  nBRG}|k\sigma\rangle.
\end{eqnarray}
Using the table of integrals in \cite{Albright1977, AiryF2004} we
obtain the RG, BRG and LV contributions to the energy spectrum of the
quantum gravitation states of unpolarized UCNs. We get
\begin{eqnarray}\label{eq:12}
 \langle \sigma k|\Phi_{\rm nRG}|k \sigma\rangle &=& \int^{\infty}_0
 dz\,\Psi^{(0)\dagger}_{k\sigma}(z)\Phi_{\rm
   nRG}\Psi^{(0)}_{k\sigma}(z) =
 \frac{2}{5}\,\frac{(E^{(0)}_k)^2}{m},\nonumber\\ \langle \uparrow
 k|\Phi^{(\rm RI)}_{\rm nBRG}|k \uparrow\rangle &=& \int^{\infty}_0
 dz\,\Psi^{(0)\dagger}_{k\uparrow}(z)\Phi^{(\rm RI)}_{\rm
   nBRG}\Psi^{(0)}_{k\uparrow}(z) = - \frac{2}{3}\,\big(k^{(\rm
   NR)}_{\phi})_n\,\frac{E^{(0)}_k}{m} - g \,\big(k^{(\rm NR)}_{\sigma
   g}\big)'_n \nonumber\\ &-& \frac{2}{3}\,m g\,E^{(0)}_k
 \Big(\big(k^{(\rm NR)}_{\sigma g pp}\big)'_n + \big(k^{(\rm
   NR)}_{\sigma g pp}\big)''_n\Big),\nonumber\\ \langle \downarrow
 k|\Phi^{(\rm RI)}_{\rm nBRG}|k \downarrow\rangle &=& \int^{\infty}_0
 dz\,\Psi^{(0)\dagger}_{k\uparrow}(z)\Phi^{(\rm RI)}_{\rm
   nBRG}\Psi^{(0)}_{k\uparrow}(z) = - \frac{2}{3}\,\big(k^{(\rm
   NR)}_{\phi})_n\,\frac{E^{(0)}_k}{m} + g \,\big(k^{(\rm NR)}_{\sigma
   g}\big)'_n \nonumber\\ &+& \frac{2}{3}\,m g\,E^{(0)}_k
 \Big(\big(k^{(\rm NR)}_{\sigma g pp}\big)'_n + \big(k^{(\rm
   NR)}_{\sigma g pp}\big)''_n\Big).
\end{eqnarray}
Since the binding energies of the quantum gravitational states of UCNs
are of a few parts of $10^{-12}\,{\rm eV}$, the RG contribution is of
order of a few parts of $10^{-33}\,{\rm eV}$ and can be
neglected. This concerns also the contributions proportional to
$\frac{2}{3}\, m g E^{(0)}_k \le 10^{-25} \,{\rm eV^3}$ for $k \le 10$
\cite{Jenke2019, Sedmik2019}.

As a result, the energy spectrum of the quantum gravitational states
of UCNs together with the BRG and LV contributions is equal to
\begin{eqnarray}\label{eq:13}
E_{k \uparrow/k \downarrow} = E^{(0)}_k - \frac{2}{3}\,\big(k^{(\rm
  NR)}_{\phi})_n\,\frac{E^{(0)}_k}{m} \mp g \,\big(k^{(\rm
  NR)}_{\sigma g}\big)'_n,
\end{eqnarray}
where $g = 2.15 \times 10^{-23}\,{\rm eV}$ \cite{Ivanov2019}. The LV
contribution, proportional to $\big(k^{(\rm NR)}_{\sigma g}\big)'_n$,
is the same for all energy level. It depends only on the neutron
spin-polarization.

According to the energy spectrum Eq.(\ref{eq:13}), for the
non-spin-flip $|k\sigma\rangle \to |k'\sigma\rangle$ and spin-flip
$|k\sigma\rangle \to |k'\sigma'\rangle$ transitions we get
\cite{Ivanov2019}
\begin{eqnarray}\label{eq:14}
\delta \nu_{k'\sigma k\sigma} &=& - \big(k^{\rm NR}_{\phi})_n\,
\frac{E^{(0)}_{k'} - E^{(0)}_k}{3\pi m},\nonumber\\ \delta
\nu_{k'\sigma' k\sigma} &=& \pm \,\frac{g}{\pi} \,\big(k^{(\rm
  NR)}_{\sigma g}\big)'_n - \big(k^{\rm NR}_{\phi})_n\,
\frac{E^{(0)}_{k'} - E^{(0)}_k}{3\pi m}.
\end{eqnarray}
 for $(\sigma = \uparrow, \sigma' = \downarrow)$ or $(\sigma =
 \downarrow, \sigma' = \uparrow)$, respectively.

For current sensitivity $\Delta E = 2\times 10^{-15}\,{\rm eV}$
\cite{Jenke2019} (see also \cite{Ivanov2019}) and for the $|1\rangle
\to |4\rangle$ transition \cite{Ivanov2019} we are able to obtain the
upper bound on the BRG contribution $\big|\big(k^{(\rm
  NR)}_{\phi})_n\big|$ and an estimate for $\big(k^{(\rm NR)}_{\sigma
  g}\big)'_n$, i.e.,
\begin{eqnarray}\label{eq:15}
\big|\big(k^{(\rm NR)}_{\phi})_n\big| < 10^{-3}\,{\rm GeV} \quad,\quad
\big(k^{(\rm NR)}_{\sigma g}\big)'_n = 0.
\end{eqnarray}
The upper bound $\big|\big(k^{(\rm NR)}_{\phi})_n\big| < 10^{-3}\,{\rm
  GeV} $ is one order of magnitude better in comparison with the
result $\big|\big(k^{(\rm NR)}_{\phi})_n\big| < 1.3 \times
10^{-2}\,{\rm GeV} $, obtained in \cite{Kostelecky2021b}. Then, our
result $\big(k^{(\rm NR)}_{\sigma g}\big)'_n = 0$ agrees well with
that by Kosteleck\'y and Li \cite{Kostelecky2021b}. The spin-flip
transitions may also admit an upper bound $\big|\big(k^{(\rm
  NR)}_{\sigma g}\big)'_n \big| < 10^8$. However, it seems
unrealistic, since the main contribution to $\big(k^{(\rm NR)}_{\sigma
  g}\big)'_n $ is caused by LV interactions \cite{Kostelecky2011b}.

It is important to emphasize that in the coefficient $(k^{\rm
  NR}_{\phi})_n$ the dominate role belongs to the BRG
interactions. According to \cite{Kostelecky2021b}, the coefficient
$(k^{\rm NR}_{\phi})_n$ has the following structure (see Table III in
Ref.\cite{Kostelecky2021b}):
\begin{eqnarray}\label{eq:16}
\big(k^{\rm NR}_{\phi})_n = 2\,(m{'}{}^{\rm L})^{ss}_n - 2\,(a^{\rm
  L})^{tss}_n + 2\,m\,(e^{\rm L}_h)^{tss}_n - 2\,m\,(c^{\rm
  L}_h)^{tss}_n+ 2\,m^2\,(m{^{(5)}_h}{}^{\rm L})^{ttss}_n -
2\,m^2\,(a{^{(5)}_h}{}^{\rm L})^{ttss}_n,
\end{eqnarray}
where the phenomenological coupling constants in the right-hand-side of
Eq.(\ref{eq:16}) are fully induced by the BRG interactions (see Table
I in Ref.\cite{Kostelecky2021b}).

The energy spectrum of the quantum gravitational states of unpolarized
UCNs, calculated by taking into account the 2-fold degeneracy of the
energy levels \cite{Ivanov2019} (see also \cite{LL1965, Davydov1965}),
is equal to
\begin{eqnarray}\label{eq:17}
E^{(\pm)}_k = E^{(0)}_k - \frac{2}{3}\,\big(k^{(\rm
  NR)}_{\phi})_n\,\frac{E^{(0)}_k}{m} \pm \Big|g \,\big(k^{(\rm
  NR)}_{\sigma g}\big)'_n \Big|.
\end{eqnarray}
Using Eq.(\ref{eq:15}) in Ref.\cite{Ivanov2019} we define the
contributions to the transition frequencies of the quantum
gravitational states of UCNs
\begin{eqnarray}\label{eq:18}
\delta \nu^{(\pm \pm)}_{k'k} &=& - \big(k^{(\rm NR)}_{\phi})_n
\,\frac{E^{(0)}_{k'} - E^{(0)}_k}{3\pi m},
\end{eqnarray}
and 
\begin{eqnarray}\label{eq:19}
\delta \nu^{(\pm \mp)}_{k'k} &=& - \big(k^{(\rm
  NR)}_{\phi})_n\,\frac{E^{(0)}_{k'} - E^{(0)}_k}{3\pi m} \pm \Big|
\frac{g}{\pi} \,\big(k^{(\rm NR)}_{\sigma g}\big)'_n\Big|.
\end{eqnarray}
One may see that the experimental analysis of the transition
frequencies between the quantum gravitational states of unpolarized
UCNs should lead to the estimates Eq.(\ref{eq:15}).

\section{Discussion}
\label{sec:Abschluss}

We have analyzed a possibility to test contributions of interactions,
induced by non-Riemann geometry beyond the standard Riemann General
Relativity and the Lorentz-invariance violation (LV), by Kosteleck\'y
and Li \cite{Kostelecky2021b}. Using the effective low-energy
potential, derived in \cite{Kostelecky2021b}, we have calculated the
contributions of beyond-Riemann gravity or the BRG contributions and
the LV contributions to the energy spectrum and transition frequencies
of the quantum gravitational states of polarized and unpolarized
UCNs. Such UCNs are used as test-particles for probes of contributions
of interactions beyond the Standard Model (SM) and Einstein's gravity
\cite{Jenke2019, Sedmik2019, Abele2021, Abele2003, Abele2009,
  Jenke2009, Abele2010, Jenke2011, Abele2011c, Abele2012, Jenke2014a,
  Jenke2014b, Schmiedmayer2015, Cronenberg2015, Abele2016, Konrad2017,
  Cronenberg2018, Ivanov2020}. We have got the following constraints
$\big|\big(k^{(\rm NR)}_{\phi})_n\big| < 10^{-3}\,{\rm GeV}$ and $
\big(k^{(\rm NR)}_{\sigma g}\big)'_n = 0$.  The upper bound
$\big|\big(k^{(\rm NR)}_{\phi})_n\big| < 10^{-3}\,{\rm GeV}$ is one
order of magnitude better in comparison with the constraint obtained
in \cite{Kostelecky2021b}. Then, from our analysis of the transition
frequencies of the quantum gravitational states of UCNs follows that $
\big(k^{(\rm NR)}_{\sigma g}\big)'_n = 0$, whereas in
\cite{Kostelecky2021b} such a value $ \big(k^{(\rm NR)}_{\sigma
  g}\big)'_n = 0$ has been imposed by assumption. It is important to
emphasize that for the experimental sensitivity $\Delta E = 2 \times
10^{-17}\,{\rm eV}$, which should be reached in the {\it q}BOUNCE
experiments in a nearest future \cite{Abele2010}, we should expect the
upper bound $\big|\big(k^{(\rm NR)}_{\phi})_n\big| < 10^{-5}\,{\rm
  GeV}$.

As has been pointed out by Kosteleck\'y and Li \cite{Kostelecky2021b},
the coefficient $(k^{\rm NR}_{\phi})_n$ should appear in the
nonrelativistic Hamilton operator in Minkowski spacetime
\cite{Kostelecky1999a} but it produces no measurable effects in that
context because it amounts to an unobservable redefinition of the zero
of energy or, equivalently, because it can be removed from the theory
via field redefinitions \cite{Kostelecky1997a, Kostelecky1998a}. The
observability of $(k^{\rm NR}_{\phi})_n$ is thus confirmed to be a
consequence of the coupling to the gravitational potential, the
presence of which restricts the applicability of field redefinitions
\cite{Kostelecky2021a}.

In the perspective of the further analysis of BRG and LV interactions
by the quantum gravitation states of UCNs we see in the use of i) the
total effective low-energy potential Eq.(\ref{eq:3}) for the
calculation of the BRG and LV contributions to the energy spectrum and
the transition frequencies of the quantum gravitational states of
UCNs, and of ii) the quantum bouncing ball experiments with a free
fall of UCNs in the gravitational field of the Earth \cite{Abele2011c,
  Abele2012}.

\section{Acknowledgements}

We are grateful to Alan Kosteleck\'y for fruitful discussions and
comments. The work of A. N. Ivanov was supported by the Austrian
``Fonds zur F\"orderung der Wissenschaftlichen Forschung'' (FWF) under
the contracts P31702-N27, P26636-N20 and P33279-N and ``Deutsche
F\"orderungsgemeinschaft'' (DFG) AB 128/5-2. The work of M. Wellenzohn
was supported by the MA 23.

\end{document}